\documentclass[lettersize,journal]{IEEEtran}

\usepackage{amsmath,amssymb,amsfonts,bm}
\usepackage{array,booktabs,multirow,makecell,longtable,tabularx,adjustbox,ragged2e}
\setcellgapes{3pt}

\usepackage{graphicx}
\usepackage[caption=false,font=normalsize,labelfont=sf,textfont=sf]{subfig}
\usepackage{stfloats}

\usepackage{algorithm,algorithmic}

\usepackage[justification=centering]{caption}
\usepackage[utf8]{inputenc}

\usepackage{cite}
\usepackage{tabularx} 
\usepackage{xcolor}
\usepackage{url}
\usepackage{verbatim}
\usepackage{ragged2e}
\usepackage{changepage}
\usepackage{pifont}

\usepackage[colorlinks=true,linkcolor=blue,citecolor=blue,urlcolor=blue]{hyperref}

\begin{document}

\title{Explainable Task-Oriented Token Communication for AI-Native 6G Networks}

\author{ Feibo Jiang, \textit{Senior Member, IEEE}, Lei Mao, Li Dong, Kezhi Wang, \textit{Senior Member, IEEE}, Cunhua Pan, \textit{Senior Member, IEEE}, Jiangzhou Wang, \textit{Fellow, IEEE} 
}

\maketitle

\begin{abstract}
The integration of Foundation Models (FMs) and wireless communications is driving the evolution of image communication from bit-accurate transmission toward task-oriented transmission. However, existing task-oriented image communication methods still face three major challenges: insufficient task-oriented Token representation, inadequate collaboration between Visual Tokens and Task Tokens, and limited interpretability of task decisions. To address these challenges, we propose an Explainable Task-Oriented Token Communication (ET-TokenCom) framework. By treating Tokens as unified units for information representation and transmission, the proposed framework constructs an end-to-end communication link that spans visual perception, wireless transmission, and task reasoning. At the transmitter, the ET-TokenCom framework extracts Visual Tokens from images to preserve low-level visual information. Meanwhile, Task Tokens generated by the FM are introduced to represent the target information and decision intent required by the current task. A Cross-Modal Attention (CMA) fusion mechanism is further designed, enabling Task Tokens to explicitly guide the selection, weighting, and transmission of Visual Tokens. At the receiver, the framework integrates Token decoding with an explainable output mechanism, where attention heatmaps are generated to highlight critical perceptual regions under different task objectives and reveal the influence of Task Tokens on the outputs. Finally, simulation results validate the effectiveness and robustness of the proposed ET-TokenCom framework.
\end{abstract}

\begin{IEEEkeywords}
Token Communication, Task-Oriented Communication, Explainable Communication, Foundation Model, Cross-Modal Attention.
\end{IEEEkeywords}

\section{Introduction}

With the rapid integration of Foundation Models (FMs) and wireless communications, image communication is evolving from pixel-level fidelity-oriented transmission toward intelligent task-aware transmission\cite{shi2023task}. Unlike conventional image communication, which primarily focuses on image reconstruction quality, FMs can assist image communication systems in identifying task-relevant objects, regions, and structural information through visual understanding, contextual modeling, task-intent interpretation, and prior reasoning. They can further support visual-importance assessment, redundant-feature compression, and robust recovery at the receiver, thereby improving image transmission efficiency and task completion capability under limited bandwidth and complex channel conditions \cite{jiang2024large}.

However, although FMs provide visual understanding and task reasoning capabilities for image communication, existing methods still struggle to achieve deep coordination among task objectives, visual information selection, and wireless transmission. Specifically, existing approaches can be broadly divided into two categories. The first category focuses on pixel-level semantic transmission, usually taking image reconstruction quality as the optimization objective. Such methods can effectively preserve low-level visual information, such as textures, edges, colors, and structural details, and are suitable for high-fidelity image recovery scenarios \cite{bourtsoulatze2019deep}. Nevertheless, they lack explicit modeling of task objectives and tend to transmit redundant information, making it difficult for limited communication resources to be concentrated on task completion. The second category focuses on task-level semantic transmission, extracting high-level information related to target objects or task intents to reduce transmission overhead and improve task utility \cite{xie2022task}. However, these methods may discard verifiable visual evidence, such as local textures, edge contours, spatial structures, and object boundaries, during compression and transmission, thereby weakening the interpretability and trustworthiness. Therefore, FM-empowered image communication still faces the following three key challenges.

\subsubsection{Insufficient Task-Oriented Token Representation}

Although FMs are capable of understanding visual content, task intents and contexts, a Token representation mechanism that can be directly controlled by task objectives has not yet been established. Existing image features are mostly generated automatically by visual encoders and are primarily designed for image compression and feature reconstruction. They lack joint characterization of task relevance, visual importance, and channel transmission requirements. In other words, current systems still find it difficult to transform the key information required by a task into selectable and transmittable Token units, leaving the transmitter without an explicit task-oriented information organization mechanism.

\subsubsection{Inadequate Collaboration Between Visual Tokens and Task Tokens} 

Even when Visual Tokens and Task Tokens can be obtained separately, establishing an effective collaborative relationship between them remains a critical challenge.  Visual Tokens mainly carry low-level visual evidence, such as local textures, edge contours, object regions, and spatial structures, whereas Task Tokens characterize the targets, regions, and intents emphasized by the current task. Without deep interaction between these two types of Tokens, the system may either transmit a large amount of task-irrelevant visual redundancy or weaken essential visual evidence by over-relying on high-level task information.

\subsubsection{Limited Interpretability of Task Decisions}

For task-oriented image communication, the system should not only provide the final task result but also explain which visual information is enhanced, which visual information is suppressed, and how Task Tokens affect the task outcome. Otherwise, even if the receiver obtains a satisfactory task output, users can hardly determine whether the result is supported by reasonable visual evidence. In high-risk scenarios such as autonomous driving and medical assistance, the lack of interpretability will directly affect trustworthy deployment, responsibility tracing, and decision-making safety.

Against this background, Token Communication (TokenCom) provides a promising implementation path for next-generation task-oriented image communication. As a fundamental information unit that connects visual perception, semantic understanding, and task reasoning in the era of FMs, Tokens can organize visual and task information across different modalities and abstraction levels within a unified representation framework\cite{dosovitskiy2020image}. Therefore, we propose an Explainable Task-Oriented Token Communication (ET-TokenCom) framework, aiming to construct a Token-level communication system that jointly supports task effectiveness and interpretability. The main contributions of this paper are summarized as follows.
\setcounter{subsubsection}{0}
\subsubsection{Dual-Branch Token Encoding}

We construct a dual-branch Token encoding structure consisting of a low-level visual perception branch and a high-level task-guided branch. The low-level branch extracts multi-scale  Visual Tokens from the original image through a visual encoder to preserve visual evidence, including local textures, edge contours, spatial structures, and regional relationships. The high-level branch leverages a pre-trained FM to extract Task Tokens, which represent the decision objectives, regional intents, and prior information emphasized by the current task. In this way, the system can simultaneously obtain transmittable Visual Tokens and Task Tokens, laying the foundation for subsequent Token selection, enhancement, and robust transmission.

\subsubsection{Cross-Modal Attention Fusion}

We propose a Cross-Modal Attention (CMA) fusion module to establish explicit associations between high-level Task Tokens and low-level Visual Tokens. Specifically, Task Tokens serve as Queries to represent task-oriented attention intents, while Visual Tokens serve as Keys and Values to carry visual evidence. Through CMA computation, the model can generate task-aware weights to enhance critical Visual Tokens, suppress redundant information, and dynamically modulate the transmission importance of Visual Tokens. This design improves Token transmission efficiency and task completion quality under limited bandwidth and noisy channel conditions.

\subsubsection{Heatmap-Based Interpretable Output}

We further embed interpretability into the output process of task-oriented TokenCom. Attention heatmaps are used to visualize the key regions attended to by the system under different task scenarios and to reveal the corresponding decision evidence. They also illustrate how Task Tokens influence the selection, enhancement, transmission, and recovery of Visual Tokens. As a result, the receiver can not only obtain task-relevant outputs but also trace the visual evidence on which the task results depend, thereby improving the transparency, trustworthiness, and decision-making safety of the system.

The remainder of this paper is organized as follows. Section II discusses the system architecture of task-oriented TokenCom and explainable TokenCom. Section III presents the proposed ET-TokenCom framework in detail. Section IV reports the experimental results. Section V discusses several open issues, and Section VI concludes this paper.

\section{Explainable and Task-Oriented TokenCom: Principles and Architectures}

\subsection{TokenCom and Related Work}

TokenCom is an emerging communication paradigm that regards tokens as the basic units for information representation, processing, and communication adaptation. Unlike conventional communication systems that perform encoding and transmission over bit streams, symbol sequences, or continuous features, TokenCom maps source information into structured, compressible, composable, and FM-understandable discrete token representations. It further enables information selection, feature processing, channel coding, and wireless transmission at the token level. Since tokens provide both information-carrying capability and interaction compatibility with FMs, TokenCom can establish a unified interface between communication systems and FMs, thereby promoting the evolution of communication from low-level data delivery toward task-oriented and intelligence-driven communication.

Recent studies have demonstrated the potential of TokenCom as a unified interface for FM-empowered communication systems from three perspectives. First, tokenized representations have been developed to map heterogeneous information, such as images, text, speech, and contextual knowledge, into a shared space, supporting the transition of communication systems from data reconstruction to task understanding and task-oriented execution \cite{qiao2025token}. Second, token-channel adaptation mechanisms have been explored to jointly optimize token importance, channel state, and resource allocation, thereby improving transmission efficiency and robustness under limited bandwidth and dynamic channel conditions \cite{ying2025joint}. Third, adaptive token selection and compact transmission mechanisms have been designed for resource-constrained and task-driven scenarios, aiming to reduce redundant tokens while preserving key information required by downstream tasks \cite{devoto2026adaptive,wei2025token}.

However, existing TokenCom studies mainly focus on token representation, compressed transmission, and robust recovery, while insufficient attention has been paid to task orientation and interpretability. Therefore, constructing a TokenCom framework that jointly supports task effectiveness, channel robustness, and decision interpretability will become an important research direction for FM-empowered next-generation intelligent communication systems.

\subsection{Task-Oriented TokenCom}

\begin{figure}[!t]
	\centering
	\includegraphics[width=\linewidth,keepaspectratio]{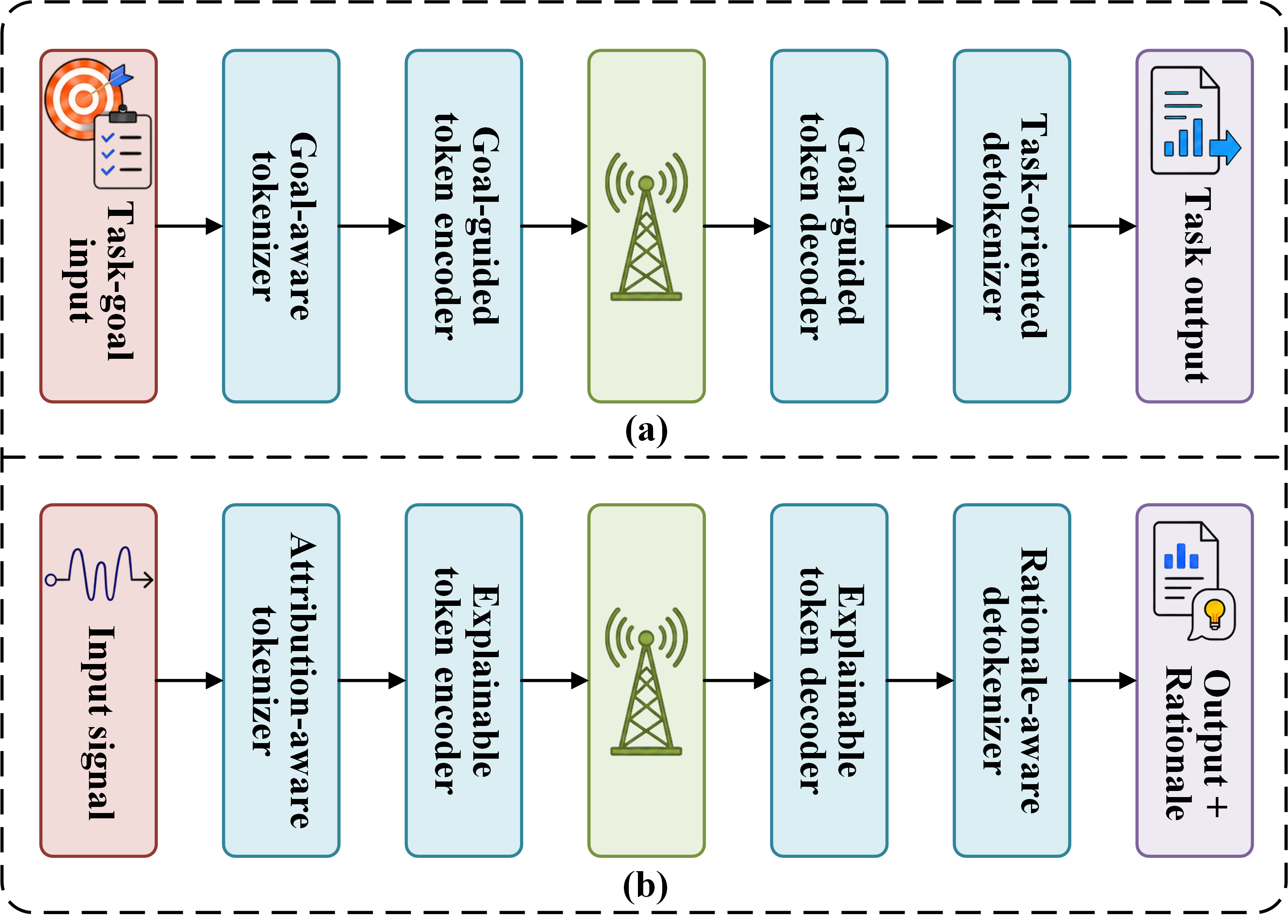}
	\caption{Architectures of Task-Oriented and Explainable TokenCom: (a) Task-Oriented TokenCom; (b) Explainable TokenCom.}
	\label{fig:task_oriented_and_explainable}
\end{figure}

Task-Oriented TokenCom is a token-level communication paradigm centered on task completion. Its basic idea is not to transmit the original data completely or uniformly compress all visual features, but to organize source information into Token representations with explicit functional meanings according to the current task requirements. Information selection, feature compression, channel transmission, and task recovery are then performed at the Token level. The system architecture of Task-Oriented TokenCom is shown in Fig. \ref{fig:task_oriented_and_explainable}(a), which consists of the following components.

\subsubsection{Task-Aware Tokenizer}

The task-aware tokenizer converts input data into structured Token representations and characterizes the relevance between different Tokens and the current task objective. Unlike a conventional tokenizer that only performs data mapping, this module further identifies high-value Tokens corresponding to key regions, target objects, or task-relevant cues, thereby providing task-aware inputs for the subsequent Token encoder.

\subsubsection{Task-Guided Token Encoder}

The task-guided Token encoder receives task-aware Token representations and maps them into coded representations suitable for wireless channel transmission. This module performs differentiated encoding according to the task contribution of each Token, allowing critical Tokens to receive stronger channel protection or more communication resources, while secondary Tokens are compressed, merged, or weakened. In this way, task-driven Token scheduling and communication resource adaptation can be achieved.

\subsubsection{Task-Guided Token Decoder}

The task-guided Token decoder recovers Token representations from noise-corrupted channel outputs at the receiver and prioritizes the reliable reconstruction of task-critical Tokens. For Tokens that are interfered with, missing, or distorted, this module can exploit contextual information, task priors, and model reasoning capabilities for completion and correction, thereby improving task completion performance and robustness under complex channel conditions.

\subsubsection{Task-Oriented Detokenizer}

The task-oriented detokenizer converts the recovered Token representations into final task-oriented outputs, such as classification labels, detection boxes, or control actions. This module does not aim to fully reconstruct the source data. Instead, it generates results for decision-making, recognition, prediction, or control according to task requirements, thereby supporting the receiver in accomplishing the intended task with limited communication overhead.

\subsection{Explainable TokenCom}

Explainable TokenCom is a new communication paradigm that explicitly incorporates interpretability mechanisms into the Token-level communication process. Its core objective is not only to use Tokens as the units of information representation, processing, and transmission to improve communication efficiency and task completion performance, but also to reveal the intrinsic relationships among task objectives, Token selection, transmission protection, and final outputs. The system architecture of Explainable TokenCom is shown in Fig. \ref{fig:task_oriented_and_explainable}(b), which consists of the following components.

\subsubsection{Attribution-Aware Tokenizer}

The attribution-aware tokenizer converts input data into structured Token representations while recording the source, attributes, and task relevance of each Token. This module further associates each Token with its corresponding content and importance rationale, such as which image region it originates from, what object or cue it corresponds to, and why it is retained or enhanced. In this way, it provides traceable inputs for subsequent encoding, transmission, and interpretability analysis.

\subsubsection{Explainable Token Encoder}

The explainable Token encoder receives the Token representations generated by the attribution-aware tokenizer and maps them into coded representations suitable for wireless channel transmission. While performing compression and channel protection, this module preserves the importance, source information, and selection rationale of Tokens. As a result, the Token filtering, encoding, and resource allocation processes at the transmitter become traceable, thereby supporting the interpretation of transmission decisions.

\subsubsection{Explainable Token Decoder}

The explainable Token decoder recovers Token representations from noise-corrupted channel outputs at the receiver and evaluates the distortion level, reliability, and task impact of different Tokens. This module not only determines whether each Token has been successfully recovered, but also analyzes which critical Tokens are affected by channel impairments and whether their distortions may influence the final task result. Therefore, it provides the basis for reliability assessment and error-source tracing.

\subsubsection{Rationale-Aware Detokenizer}

The rationale-aware detokenizer converts the recovered Token representations into final task outputs while simultaneously generating explanatory information that supports these outputs. This module not only produces task results, such as classification, detection, control, or question answering, but also reveals which Tokens contribute more significantly to the final decision. In this way, it establishes a traceable explanation chain from Token selection and channel transmission to task decision-making.

\section{Explainable Task-Oriented TokenCom Framework}

\begin{figure*}[!t]
	\centering
	\includegraphics[
	width=\textwidth,
	keepaspectratio
	]{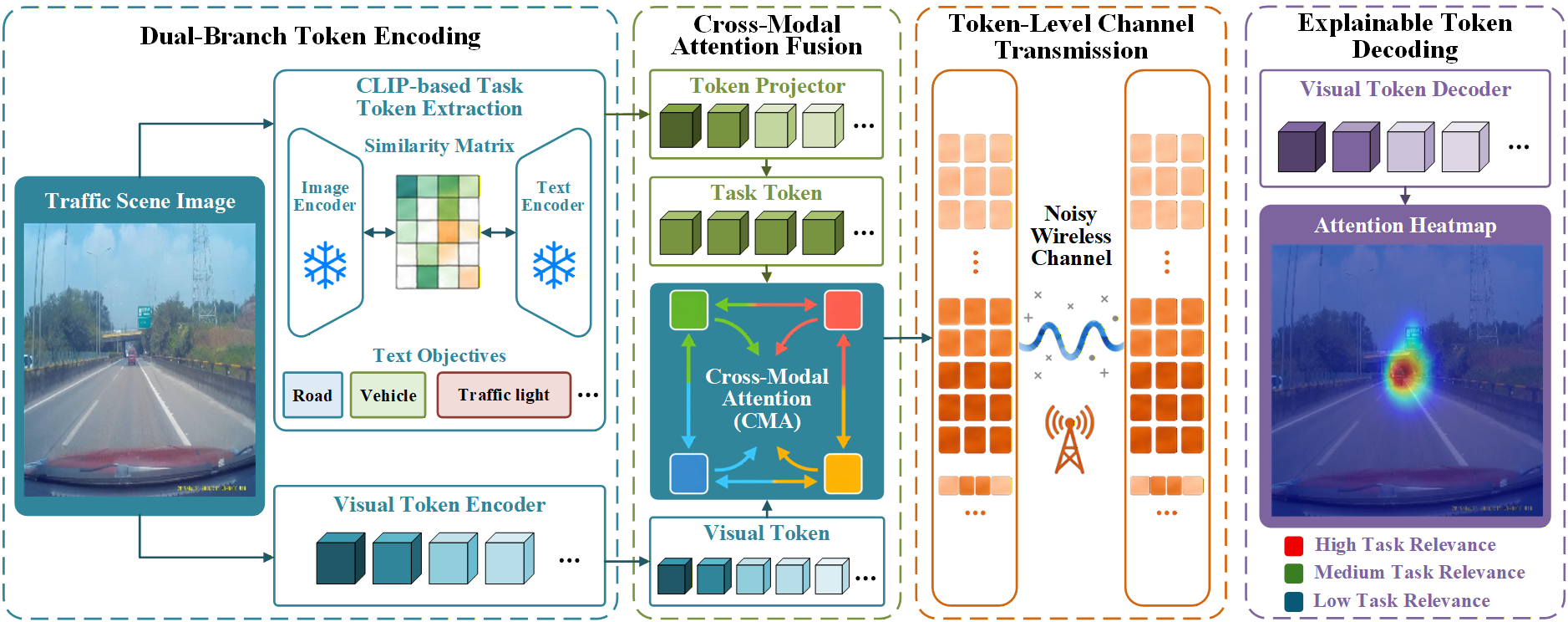}
	\caption{The Proposed Explainable Task-Oriented TokenCom Framework.}
	\label{fig:overall_framework}
\end{figure*}

The proposed ET-TokenCom framework consists of four stages: dual-branch Token encoding, CMA fusion, Token-level channel transmission, and explainable Token decoding, as shown in Fig. \ref{fig:overall_framework}.

\subsection{Overall Workflow}

\subsubsection{Dual-Branch Token Encoding}

This module first converts the input image into basic Tokens through initial tokenization, which are then fed into a low-level visual perception branch and a high-level task-guided branch for encoding. The low-level branch extracts multi-scale Visual Tokens through a visual encoder to represent fine-grained visual evidence, such as local textures, edge structures, object contours, and spatial relationships. The high-level branch leverages a pre-trained FM to generate Task Tokens, which characterize the targets, regions, and evidential cues that should be emphasized in the current task.

\subsubsection{Cross-Modal Attention Fusion}

After obtaining low-level Visual Tokens and high-level Task Tokens, the system needs to establish a guidance relationship from Task Tokens to Visual Tokens. To this end, we design a CMA fusion module, where high-level Task Tokens serve as task-guided signals and low-level Visual Tokens serve as carriers of visual evidence. Through attention weights, the module performs task-driven modulation of Visual Tokens, enhancing those highly relevant to the task objective while appropriately suppressing redundant or weakly related Tokens. Furthermore, the fused task-guided Visual Tokens are compressed and mapped into coded representations suitable for wireless channel transmission, thereby enabling robust delivery of critical visual information.

\subsubsection{Token-Level Channel Transmission}

After CMA fusion, the Tokens are
then transmitted through the wireless channel. During transmission, channel noise introduces perturbations to the Token representations. Since the encoding and fusion stages have already enhanced task-critical Visual Tokens and suppressed redundant visual information according to the task objective, the Token representations entering the channel exhibit higher task relevance and a more compact information distribution.

\subsubsection{Explainable Token Decoding}

At the receiver, the Token decoder progressively reconstructs the spatial response structure from noise-perturbed Token representations and generates task-related attention heatmaps. These heatmaps visually indicate the key regions attended to by the system during task execution, such as target objects, important structures, and interaction regions. In this way, they reveal the intrinsic relationships among task objectives, critical visual regions, and final output results \cite{10679559}.

\subsection{Token Encoder--Decoder Module}

\begin{figure}[!t]
	\centering
	\includegraphics[width=\linewidth,keepaspectratio]{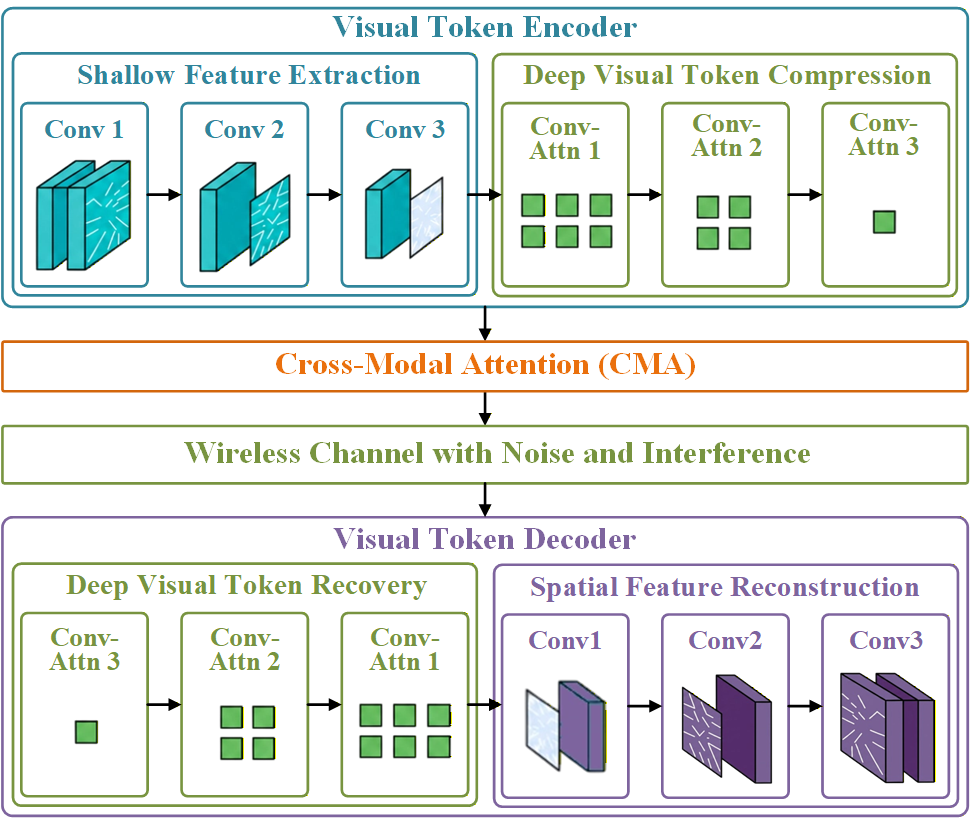}
	\caption{Token Encoder and Decoder.}
	\label{fig:bottom-up}
\end{figure}

The Token encoder--decoder module is designed to perform compressed representation of Visual Tokens, channel-adaptive transmission, and spatial feature reconstruction. Its overall architecture is shown in Fig. \ref{fig:bottom-up}. The encoder--decoder module mainly consists of Conv modules and the proposed Conv-Attn modules. Conv-Attn first employs an Inception-style multi-branch convolutional structure to perform spatial projection on intermediate feature maps, thereby extracting multi-scale local features with different receptive fields. The features from each branch are then fed into a self-attention module to model long-range dependencies among spatial regions. Finally, the outputs of all branches are concatenated and reshaped into spatial feature maps, allowing the compact Token representation to preserve both local details and global contextual information. Based on this structure, the detailed design of the Token encoder--decoder is introduced as follows.


\subsubsection{Shallow Feature Extraction}

The input image first passes through three convolutional modules to extract high-resolution low-level visual information, such as edges, textures, color variations, and local structures. This stage focuses on preserving verifiable local visual evidence in the image, thereby providing a stable feature basis for subsequent deep Token compression. Unlike high-level Task Tokens that emphasize task objectives, shallow Visual Tokens are mainly derived from the image content itself and capture fine-grained visual cues that support task execution.

\subsubsection{Deep Visual Token Compression}

After shallow feature extraction, the Token encoder further performs attention-based encoding on visual features through three Conv-Attn modules. As the network depth increases, the resolution of feature maps gradually decreases, while the receptive field and contextual representation capability are progressively enhanced. This stage organizes local visual features into more compact deep Visual Tokens, which preserve spatial structures and local visual evidence while serving as suitable information units for subsequent attention modulation and wireless transmission.

\subsubsection{Deep Visual Token Recovery}

At the receiver, the task-oriented Tokens transmitted over the wireless channel are fed into the Token decoder. This process can be regarded as the inverse operation of deep Visual Token compression. Specifically, three Conv-Attn modules are used to progressively expand the compact Token representations into higher-resolution spatial features. In contrast to the transmitter, which focuses on visual compression, the decoder emphasizes the recovery of spatial structures and task-relevant visual evidence retained in the Tokens, thereby providing the basis for subsequent spatial feature reconstruction.

\subsubsection{Spatial Feature Reconstruction}

After deep Visual Token recovery, the decoder further maps the features into spatial response maps through three convolutional modules. This process corresponds to the reverse projection of shallow visual feature extraction, aiming to remap the compressed Token information back into the image space. It should be emphasized that this module does not aim to reconstruct the original image as the final objective. Instead, it generates task-related spatial attention responses to characterize the contribution of different image regions to the final task output.

\subsection{Task Token Extraction}

As shown in Fig. \ref{fig:overall_framework}, we employ CLIP, a pre-trained vision-language FM, to extract Task Tokens from the input image \cite{radford2021learning}.

\subsubsection{Task Token Generation}

CLIP performs Token encoding on the input image through its image-text alignment capability and directly generates Task Tokens. These Tokens are used to represent the decision objectives, regional intents, and prior information emphasized by the current task. Different from Visual Tokens, which mainly focus on local textures and spatial details, Task Tokens place greater emphasis on task-relevant global information and provide guidance for subsequent Visual Token selection and enhancement.

\subsubsection{Shared Token Projection}

Since Task Tokens are derived from the pre-trained vision-language FM, whereas Visual Tokens are generated by the visual encoder, their feature dimensions are not fully consistent. Therefore, we introduce a Token projection module to map Task Tokens into a shared space compatible with Visual Tokens, thereby providing a unified representation for subsequent CMA fusion.

\subsection{Cross-Modal Attention Fusion Module}

\begin{figure}[!t]
	\centering
	\includegraphics[
	width=\linewidth,		
	keepaspectratio
	]{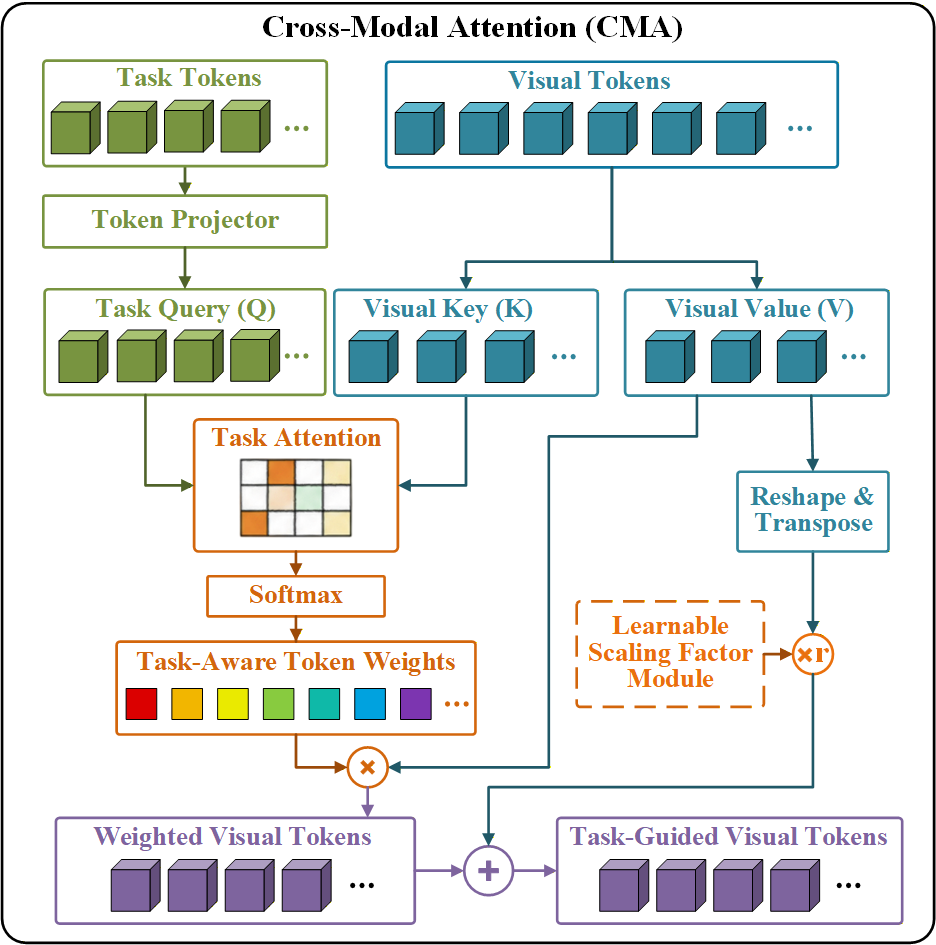}
	\caption{Cross-Modal Attention.}
	\label{fig:cma}
\end{figure}

As shown in Fig. \ref{fig:cma}, the CMA module consists of two main components: task-aware weight generation and task-guided Visual Token fusion.

\subsubsection{Task-Aware Weight Generation}

\textcolor{black}{In the shared representation space, Task Tokens are mapped into task Queries, while Visual Tokens are mapped into visual Keys. The module first computes cross-modal attention between task Queries and visual Keys, and then applies Softmax normalization to generate task-aware Token weights.}

\subsubsection{Task-Guided Visual Token Fusion}

\textcolor{black}{Visual Tokens are further mapped into visual Values, which serve as the main representations of low-level visual evidence. The task-aware Token weights are then applied to the visual Values to obtain weighted Visual Tokens. Subsequently, the CMA module modulates the visual Values using learnable scaling factors and fuses them with the weighted results, ultimately generating task-guided Visual Tokens.}

\section{Simulation Results}

\subsection{Experimental Settings}

This experiment simulates an image Token communication task in an intelligent Internet of Vehicles scenario, where the receiver aims to recover attention heatmaps for safety-critical driving tasks. In the proposed ET-TokenCom framework, the Task Token extraction module is implemented using CLIP ViT-B/32. The Visual Token encoder and decoder are constructed with an autoencoder-like architecture, each consisting of three convolutional layers and three Conv-Attn layers. The Token projection layer is implemented using a linear layer. The wireless channel model follows a setting similar to that in \cite{jiang2024large}. The TrafficGaze dataset is used for training \cite{zhao2025salmamba}.

During training, the Adam optimizer is adopted, with the initial learning rate set to \(1 \times 10^{-3}\), the weight decay coefficient set to \(1 \times 10^{-4}\), the number of training epochs set to 100, and the batch size set to 64. The experimental loss function consists of saliency prediction loss and feature reconstruction loss, where Binary Cross-Entropy (BCE) loss is used for saliency prediction and Mean Squared Error (MSE) loss is used for feature reconstruction.

The experiments are implemented in Python 3.10.19 using the PyTorch 2.2.0 with CUDA 12.4. The hardware platform consists of an Intel Xeon CPU operating at 2.6 GHz with approximately 1007 GB of memory and an NVIDIA A800-SXM4-80GB GPU.

\subsection{Ablation Analysis}

This experiment mainly evaluates the impact of Task Tokens and cross-modal fusion on explainable task outputs in the proposed framework. Here, ``w/o Task Tokens'' denotes the model that does not rely on Task Tokens or CMA fusion, while ``Ours'' denotes the proposed method. Visual heatmap comparison is adopted as the evaluation approach to intuitively illustrate the spatial regions attended to by different models in driving scenarios and to analyze whether they can accurately capture key information related to driving safety.

\begin{figure}[!t]
		\centering
		\includegraphics[
		width=\linewidth,
		keepaspectratio
		]{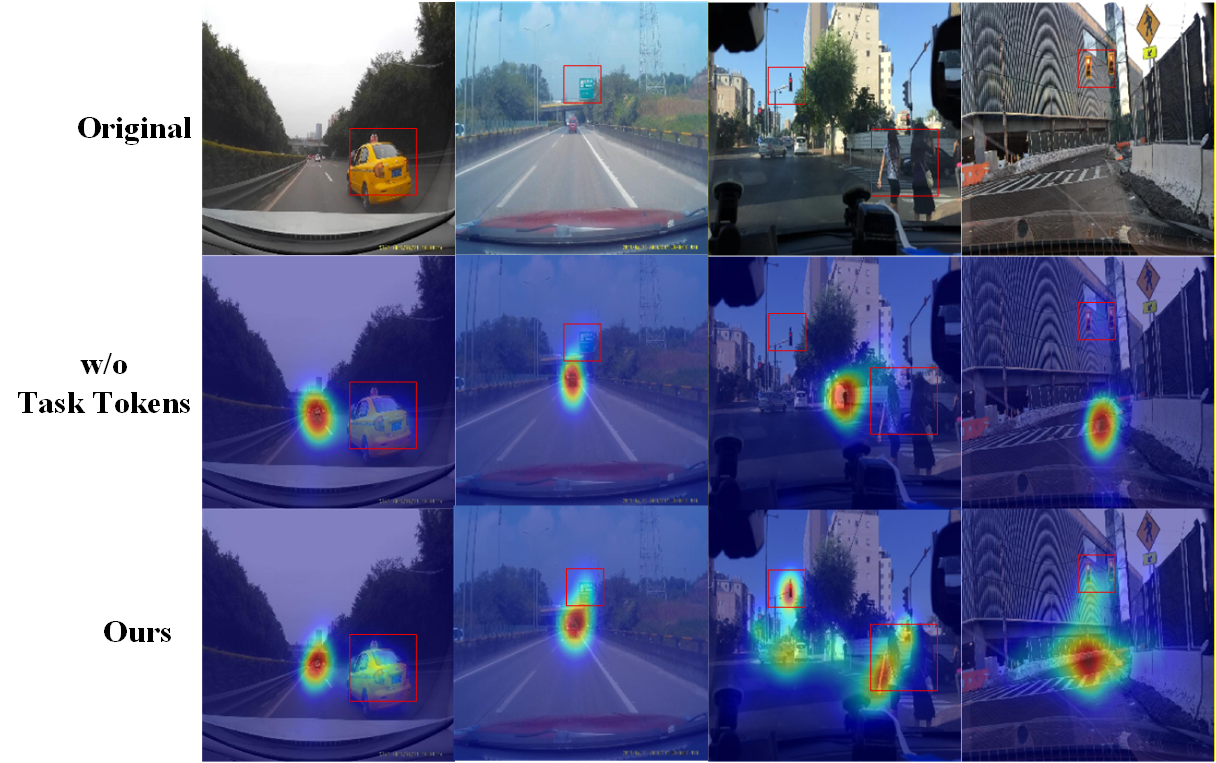}
		\caption{Ablation Analysis of Task Token Guidance.}
		\label{fig:ablation}
\end{figure}

As shown in Fig. \ref{fig:ablation}, the model relying only on low-level visual features mainly focuses on the most direct targets in the image, such as the frontal road region, but pays insufficient attention to regions semantically related to driving safety, such as potential risk objects and nearby traffic signs. In contrast, under the guidance of high-level Task Tokens, the complete model can not only focus on visually salient targets but also further capture semantically associated regions relevant to safe driving decisions. The generated heatmaps are more consistent with the real driver attention distribution. This result indicates that Task Token-guided CMA fusion can enhance the model's perception of task-relevant semantic targets.

\subsection{Robustness Analysis}

This experiment is mainly conducted to verify the task completion performance and robustness of the proposed framework under different Signal-to-Noise Ratio (SNR) conditions. We select SCOUT \cite{kotseruba2024scout+} and CDNN \cite{deng2019drivers} as baseline methods and uniformly introduce channel interference into all methods to evaluate their performance degradation and task recovery capability in wireless noise environments. Similarity metric (SIM) is adopted as the primary metric \cite{borji2012quantitative}, which measures the similarity between the predicted attention heatmap and the real driver attention distribution. A higher SIM value indicates that the model can more accurately characterize the driver attention regions in complex traffic scenarios.

\begin{figure}[!t]
		\centering
		\includegraphics[
		width=\linewidth,
		keepaspectratio
		]{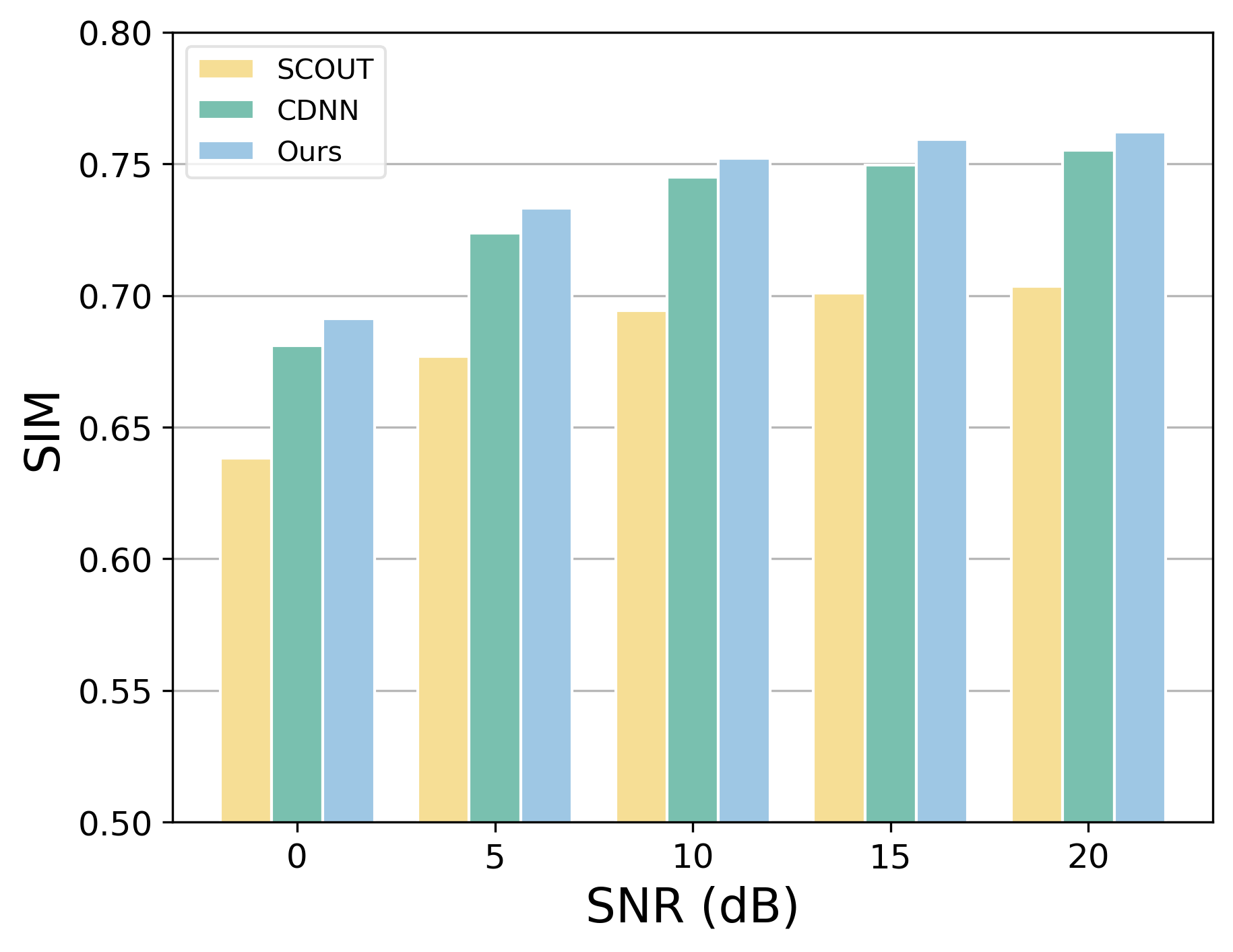}
		\caption{Robustness Analysis under Different SNR Conditions.}
		\label{fig:sim}
\end{figure}

As shown in Fig. \ref{fig:sim}, the proposed method achieves higher SIM values than the baseline methods under all SNR conditions. This demonstrates that the proposed framework can not only generate accurate attention distributions under favorable channel conditions but also maintain strong task robustness under severe noise interference. The main reason is that the high-level Task Tokens and CMA fusion mechanism can enhance critical Visual Tokens related to driving safety before transmission and suppress redundant background information, making the Token representations entering the channel more compact and task-relevant.

\section{Open Issues}

\subsubsection{Task-Adaptive Token Representation}

In this paper, high-level Task Tokens are used to guide the selection and enhancement of low-level Visual Tokens. However, current Task Tokens are still mainly generated based on pre-trained FMs or fixed task priors. Different image communication tasks may have distinct requirements in terms of Token granularity, content type, and representation level. Therefore, future research should develop task-adaptive Token representation mechanisms, enabling Task Tokens to dynamically adjust according to task objectives, scenario variations, and receiver-side requirements.

\subsubsection{Task-Channel Collaborative Token Transmission}

This paper mainly enhances or suppresses Visual Tokens according to their task relevance. However, in practical wireless channels, SNR, bandwidth, interference, and time-varying fading directly affect the transmission reliability of different Tokens. Considering only task importance may cause critical Tokens to be severely corrupted under adverse channel conditions. Therefore, future research should jointly model task importance and channel reliability to develop collaborative Token selection, protection, and transmission mechanisms.

\subsubsection{Verifiable Token-Level Explanation Mechanism}

This paper uses attention heatmaps to illustrate how Task Tokens modulate Visual Tokens, thereby reflecting the importance of different Visual Tokens to the current task output. However, heatmaps mainly provide visualization results of spatial importance and still have limited ability to quantify the necessity and stability of critical Tokens for the final decision. In the future, Token occlusion, counterfactual replacement, and causal intervention can be introduced to advance Explainable TokenCom from ``visualized explanation'' toward ``verifiable explanation''.

\section{Conclusion}

In this paper, we investigated the issues of task effectiveness, channel robustness, and decision interpretability in task-oriented image TokenCom, and proposed an ET-TokenCom framework. The proposed framework regards Tokens as unified units for information representation and processing. By integrating low-level Visual Token representation, high-level Task Token guidance, and CMA fusion, the framework effectively enhances task-relevant Visual Tokens, suppresses redundant information, and enables robust transmission over noisy channels. Meanwhile, the framework visualizes the modulation effect of Task Tokens on the response distribution of Visual Tokens through attention heatmaps, revealing the intrinsic relationship between critical visual regions and final task outputs. This improves the transparency and trustworthiness of the system. Overall, this work promotes the evolution of TokenCom from an efficient transmission paradigm toward a task-effective, channel-robust, and decision-trustworthy communication paradigm.

\bibliographystyle{IEEEtran}
\bibliography{references}

@article{deng2019drivers,
	title={How do drivers allocate their potential attention? Driving fixation prediction via convolutional neural networks},
	author={Deng, Tao and Yan, Hongmei and Qin, Long and Ngo, Thuyen and Manjunath, BS},
	journal={IEEE Transactions on Intelligent Transportation Systems},
	volume={21},
	number={5},
	pages={2146--2154},
	year={2019},
	publisher={IEEE}
}

@INPROCEEDINGS{kotseruba2024scout+,
  author={Kotseruba, Iuliia and Tsotsos, John K.},
  booktitle={2024 IEEE Intelligent Vehicles Symposium (IV)}, 
  title={SCOUT+: Towards Practical Task-Driven Drivers' Gaze Prediction}, 
  year={2024},
  volume={},
  number={},
  pages={1927-1932},
  doi={10.1109/IV55156.2024.10588743}}

@article{borji2012quantitative,
	title={Quantitative analysis of human-model agreement in visual saliency modeling: A comparative study},
	author={Borji, Ali and Sihite, Dicky N and Itti, Laurent},
	journal={IEEE Transactions on Image Processing},
	volume={22},
	number={1},
	pages={55--69},
	year={2012},
	publisher={IEEE}
}

@article{jiang2024large,
	title={Large AI model-based semantic communications},
	author={Jiang, Feibo and Peng, Yubo and Dong, Li and Wang, Kezhi and Yang, Kun and Pan, Cunhua and You, Xiaohu},
	journal={IEEE Wireless Communications},
	volume={31},
	number={3},
	pages={68--75},
	year={2024},
	publisher={IEEE}
}

@article{bourtsoulatze2019deep,
	title={Deep joint source-channel coding for wireless image transmission},
	author={Bourtsoulatze, Eirina and Kurka, David Burth and G{\"u}nd{\"u}z, Deniz},
	journal={IEEE Transactions on Cognitive Communications and Networking},
	volume={5},
	number={3},
	pages={567--579},
	year={2019},
	publisher={IEEE}
}

@article{xie2022task,
	title={Task-oriented multi-user semantic communications},
	author={Xie, Huiqiang and Qin, Zhijin and Tao, Xiaoming and Letaief, Khaled B},
	journal={IEEE Journal on Selected Areas in Communications},
	volume={40},
	number={9},
	pages={2584--2597},
	year={2022},
	publisher={IEEE}
}

@inproceedings{radford2021learning,
	title={Learning transferable visual models from natural language supervision},
	author={Radford, Alec and Kim, Jong Wook and Hallacy, Chris and Ramesh, Aditya and Goh, Gabriel and Agarwal, Sandhini and Sastry, Girish and Askell, Amanda and Mishkin, Pamela and Clark, Jack and others},
	booktitle={International conference on machine learning},
	pages={8748--8763},
	year={2021},
	organization={PmLR}
}

@article{qiao2025token,
	title={Token communications: A large model-driven framework for cross-modal context-aware semantic communications},
	author={Qiao, Li and Mashhadi, Mahdi Boloursaz and Gao, Zhen and Tafazolli, Rahim and Bennis, Mehdi and Niyato, Dusit},
	journal={IEEE Wireless Communications},
	volume={32},
	number={5},
	pages={80--88},
	year={2025},
	publisher={IEEE}
}

@article{ying2025joint,
	title={Joint Semantic-Channel Coding and Modulation for Token Communications},
	author={Ying, Jingkai and Qin, Zhijin and Feng, Yulong and Wang, Liejun and Tao, Xiaoming},
	journal={IEEE Transactions on Wireless Communications},
	volume={25},
	pages={8179--8193},
	year={2025},
	publisher={IEEE}
}

@article{devoto2026adaptive,
	title={Adaptive semantic token communication for transformer-based edge inference},
	author={Devoto, Alessio and Pomponi, Jary and Merluzzi, Mattia and Di Lorenzo, Paolo and Scardapane, Simone},
	journal={IEEE Transactions on Machine Learning in Communications and Networking},
	year={2026},
	publisher={IEEE}
}

@article{wei2025token,
	title={Token Communication in the Era of Large Models: An Information Bottleneck-Based Approach},
	author={Wei, Hao and Ni, Wanli and Wang, Wen and Xu, Wenjun and Niyato, Dusit and Zhang, Ping},
	journal={IEEE Wireless Communications Letters},
	year={2025},
	publisher={IEEE}
}

@article{zhao2025salmamba, 
	title={SalM²: An Extremely Lightweight Saliency Mamba Model for Real-Time Cognitive Awareness of Driver Attention}, 
	volume={39}, 
	DOI={10.1609/aaai.v39i2.32157},  
	number={2},
	journal={Proceedings of the AAAI Conference on Artificial Intelligence}, 
	author={Zhao, Chunyu and Mu, Wentao and Zhou, Xian and Liu, Wenbo and Yan, Fei and Deng, Tao}, 
	year={2025}, 
	month={Apr.}, 
	pages={1647-1655} 
}

@article{shi2023task,
	title={Task-oriented communications for 6G: Vision, principles, and technologies},
	author={Shi, Yuanming and Zhou, Yong and Wen, Dingzhu and Wu, Youlong and Jiang, Chunxiao and Letaief, Khaled B},
	journal={IEEE Wireless Communications},
	volume={30},
	number={3},
	pages={78--85},
	year={2023},
	publisher={IEEE}
}

@article{dosovitskiy2020image,
	title={An image is worth 16x16 words: Transformers for image recognition at scale},
	author={Dosovitskiy, Alexey and Beyer, Lucas and Kolesnikov, Alexander and Weissenborn, Dirk and Zhai, Xiaohua and Unterthiner, Thomas and Dehghani, Mostafa and Minderer, Matthias and Heigold, Georg and Gelly, Sylvain and others},
	journal={arXiv preprint arXiv:2010.11929},
	year={2020}
}

@ARTICLE{10679559,
	author={Dong, Li and Peng, Yubo and Jiang, Feibo and Wang, Kezhi and Yang, Kun},
	journal={IEEE Transactions on Industrial Informatics}, 
	title={Explainable Semantic Federated Learning Enabled Industrial Edge Network for Fire Surveillance}, 
	year={2024},
	volume={20},
	number={12},
	pages={14053-14061},
	doi={10.1109/TII.2024.3441626}}
\section*{Biographies}

\textbf{Feibo Jiang} (jiangfb@hunnu.edu.cn) is currently an Associate Professor at Hunan Normal University, China.

\textbf{Lei Mao} (maolei@hunnu.edu.cn) is currently pursuing the master’s degree at Hunan Normal University, China.

\textbf{Li Dong} (Dlj2017@hunnu.edu.cn) is currently a Professor at Hunan University of Technology and Business, China.

\textbf{Kezhi Wang} (Kezhi.Wang@brunel.ac.uk) is a Professor with the Department of Computer Science, Brunel University London, U.K.

\textbf{Cunhua Pan} (cpan@seu.edu.cn) is a Full Professor in Southeast University, China.

\textbf{Jiangzhou Wang} (j.z.wang@seu.edu.cn) is a Full Professor in Southeast University, China.

\end{document}